\documentclass[11pt,twoside]{article}
\usepackage{asp2010}

\resetcounters

\begin{document}

\title{The Design and Implementation of the Akamai Maui Short Course}
\author{Ryan Montgomery$^1$, Dave Harrington$^2$, Sarah Sonnett$^2$,
  Mark Pitts$^2$, Isar Mostafanezhad$^3$, Mike Foley$^4$, Eddie Laag$^5$,
  and Lisa Hunter$^6$
  \affil{$^1$Department~of~Astronomy~and~Astrophysics, 
    University~of~California~at~Santa~Cruz,
    1156~High~St., Santa`Cruz, CA 95064}
  \affil{$^2$Institute~for~Astronomy, 
    University~of~Hawai`i~at~Manoa, 
    2680~Woodlawn~Dr., Honolulu, HI 96822}
  \affil{$^3$Department~of~Electrical~Engineering,
    University~of~Hawai`i~at~Manoa, 
    2540~Dole~St., Honolulu, HI 96822}
  \affil{$^4$Hawai`i~Natural~Energy~Institute,
    University~of~Hawai`i~at~Manoa,
    1680~East-West~Rd., Honolulu, HI 96822}
  \affil{$^5$Department~of~Remote~Sensing,
    The~Aerospace~Corporation, 
    2310~E.~El~Segundo~Blvd, El~Segundo, CA 90245}
  \affil{$^6$Institute~for~Scientist~and~Engineer~Educators,
    University~of~California~at~Santa~Cruz, 
    1156~High~St., Santa~Cruz, CA 95064}
}

\begin{abstract}
We describe the design and elements of implementation of the Akamai Maui Short Course (AMSC). The AMSC contains four full inquiry activities each of which builds on those previous: Camera Obscura and Sun Shadows, Lenses and Refraction, Color and Light, and the Adaptive Optics Demonstrator. In addition we describe the workings of two additional strands: 1) Communication, and 2) Science, Technology and Society. We also discuss our assessment methods and our results.
\end{abstract}

\section{Introduction}
The Akamai Maui Short Course (AMSC) is a component of the larger Akamai Internship Program\footnote{Information on the Akamai Internship Program can be found at:\\ http://kopiko.ifa.hawaii.edu/akamai/internships/index.php}, which places college students from Hawaii in the high tech industry on Maui. The goal of the program is to retain and advance students in science and engineering fields, and to create a local technical workforce. The AMSC is the first week of the overall eight-week program, and is designed to prepare students to be successful in their coming internships. The Akamai Internship Program accepts students from diverse backgrounds, and many who have had limited exposure to research and/or the technical workplace. The program is designed to support students earlier in their college education, such as students in their freshmen and sophomore years at community colleges, and AMSC is a key element in helping students to be successful. The course includes four inquiry activities in which students must work in a much more self-directed way and figure out their own way to solve problems. They work together in teams to investigate curious phenomena, and are continuously called upon to express their questions, their ideas, and their explanations. 

In order to best prepare the students for their internships the short course's activities are tailored to the specific needs of the internship host sites. In the next section we enumerate several goals we have for our students which are largely determined by the abilities considered desirable in an intern by a survey of the host-site employers. In order to facilitate this connection between the interns and their host companies our short course also has activities where the student studies their host site/institution, its expertise and its limitations. Also, the students meet with a representative from their host site, discuss the upcoming collaboration and deliver a brief presentation outlining the specifics of their upcoming role in their host company. 

\section{The Need for a Diverse Local Workforce in Hawai`i}
Participants in the Akamai Internship Program all have ties to Hawaii, and are college level students in science, technology, engineering and math (STEM) fields. The aims of the program include increasing the diversity of STEM fields in Hawaii, and building a local workforce, thus the program engages in intensive and targeted recruiting to ensure that students from underrepresented groups are included. To date, 144 students have been in the program, with demographics as follows: 34\% women; 22\% Native Hawaiian/Pacific Islanders (NHPI); 49\% NHPI and other underrepresented minorities; 62\% born in Hawaii; 37\% started at a community college. It should be noted that these internships are primarily in the areas of electrical engineering, electronics, computer engineering, physics, mechanical engineering, and other fields where participation by women is still dishearteningly low. For example, women earned 12\% of Associates degrees in electrical engineering in 2007 \citep{SEI}.  At the bachelor's degree level women earned a small fraction of the degrees awarded in 2007 in engineering, computer sciences, and physics (19\%, 19\%, and 21\%, respectively). Thus the inclusion of 34\% women in the program is a relatively high number. Native Hawaiians are significantly underrepresented in the technical fields served by Akamai -- telescopes and the related spin-off industries that support astronomical research. 

As stated earlier, a major goal of the Akamai Program is to develop a local technical workforce, which includes the full diversity of the people of Hawaii. Many young people in the community would like to remain in their home state, and be part of the growing science and technology industry in Hawaii, and many high tech employers strive to hire local talent. However, there are currently limited pathways for college students, so there is a disconnect between two needs that are clearly complementary. The Akamai Program is aimed at filling this gap, currently maintaining a retention rate of 83\% of its 144 participants\footnote{http://kopiko.ifa.hawaii.edu/akamai/internships/outcomes.html}, and the AMSC is considered to be a crucial part of the overall experience.

\section{Goals for Learners}
Students, as a result of this Short Course, will gain understanding and facility in a variety of engineering and scientific skills and techniques that will help them prepare for their summer internship roles. Thus, an important goal for the course is to provide a solid foundation in relevant process, attitudinal and organizational skill sets. In addition, our instruction will emphasize content taken from the main areas of research conducted by the Center for Adaptive Optics (CfAO). Because, in essence, CfAO researches telescope technologies, we choose the study of telescopes and the underlying physics as our primary content.

\subsection{Science and Engineering Processes:}
Through the teaching of the telescope content, we would like students to learn the following essential process goals:
\begin{enumerate}
  \vspace{-2mm}
\item Ability to break a complex system into simple interacting parts.
  \vspace{-2mm}
\item Defining ``good enough'' solutions and performance criteria.
  \vspace{-2mm}
\item An ability to form and express hypotheses.
  \vspace{-2mm}
\item Basic principles of experimental and engineering design, including variable control and troubleshooting.
\end{enumerate}

\subsection{Content:}
The following are the elements of content that we consider important and relevant:
\begin{enumerate}
\item Optical and physical basics of light, lenses and telescopes:
  \vspace{-2mm}
  \begin{enumerate}
    \item Pinhole cameras and the basics of apertures;
    \item Refraction, simple aberrations, and the role of lenses;
    \item Additive and subtractive mixing of colors of light;
    \item The effect of the atmosphere (complex aberration);
    \item Methods of correcting atmospheric aberration with adaptive optics.
    \end{enumerate}
\item An overview of the two major telescope design philosophies, and specific examples of designs.
  \vspace{-2mm}
  \item An overview of the many industrial, scientific, and military uses of optics in remote sensing with a focus on Maui applications.
\end{enumerate}

\subsection{Communication and Professional Skills:}
\begin{enumerate}
\item An understanding of how to integrate and interact in a science or engineering based workspace.
  \vspace{-2mm}
\item An ability to explain results in written, visual, and oral formats. This includes the creation of appropriately structured abstracts and summaries.
  \vspace{-2mm}
\item An understanding of career opportunities in the Maui high tech industry.
\end{enumerate}

\section{Short Course Description:}
\subsection{Overview}
For reference, we include the overall short course schedule in Table 1.

\begin{table}[ht]
\begin{center}
\caption{Short Course Timeline}
\smallskip
\begin{tabular}{lcc}
\tableline
{\bf Activity} & {\bf Day \#} & {\bf Approx.\ Duration} \\
\tableline
\tableline
\noalign{\smallskip}
Opening & 0 & 1 hours \\
\noalign{\smallskip}
\noalign{\smallskip}
Introductions & 1 & 0.5 hours \\
Pre-knowledge Survey & 1 & 0.5 hours \\
Maui High Tech Investigations & 1 & 2 hours \\
Science, Technology and Society & 1/5 & 1 hours \\
Camera Obscura and Sun Shadows Inquiry & 1 & 3.25 hours \\
\noalign{\smallskip}
\noalign{\smallskip}
Anatomy of an Abstract & 2 & 1 hours \\
Lenses and Refraction Inquiry & 2 & 3.75 hours \\
Build a Telescope Activity & 2 & 1.25 hours \\
Telescope Design/Configuration Lecture & 2 & 0.5 hours \\
\noalign{\smallskip}
\noalign{\smallskip}
Peer Review of Abstracts & 3 & 0.75 hours \\
Color and Light Inquiry & 3 & 4.75 hours \\
Handheld Spectroscope Activity & 3 & 1 hours \\
Solar Astronomy Lecture & 3 & 0.75 hours \\
\noalign{\smallskip}
\noalign{\smallskip}
Haleakala Summit/Telescope Tour & 4 & 5 hours \\
\noalign{\smallskip}
\noalign{\smallskip}
Adaptive Optics Workstation Activity & 5 & 3.75 hours \\
Strategies for successfully working with mentors & 5 & 1.75 hours \\
Post-knowledge Survey & 5 & 0.5 hours \\
Closing & 5 & 0.75 hours \\
\tableline
\end{tabular}
\end{center}
\end{table}

In the following sections we will describe the four inquiries taught over the short course and then the communication component of the short course. The AMSC instructors have all participated in the PDP \citep{PDPdesc}, which prepares the instructors for designing inquiry activities and uses the inquiry model developed by the Institute for Inquiry\footnote{http://www.instituteforinquiry.org/}. A typical inquiry activity is composed of three main phases:
\begin{description}
\item[Starters --] Learners notice puzzling phenomena, which stimulate question raising. Often there are three or four stations with different phenomena all related to the same content area.
\item[Investigations --] Small teams of 2-3 learners design a way to come to their own understanding of phenomena seen in the starters.
\item[Meaning making --] Learners share what they learned by explaining their findings to the whole group. Instructor then synthesizes what the group learned collectively, covers anything missed, and summarizes.
\end{description}

\subsection{Camera Obscura and Sun Shadows}
\begin{table}[ht]
\begin{center}
\caption{Camera Obscura \& Sun Shadows Timeline}
\smallskip
\begin{tabular}{lc}
\tableline
{\bf Activity Component} & {\bf Time Allotted} \\
\tableline
\tableline
\noalign{\smallskip}
Context and Introduction & 15 min \\
Starters & 15 min \\
Camera Activities (stations 1 \& 2) & 45 min \\
Sun-shadow Activity (station 3) & 30 min \\
Presentation Preparation & 30 min \\
Poster Presentations & 30 min \\
Synthesis & 30 min \\
\tableline
\end{tabular}
\end{center}
\end{table}

The first inquiry activity of the Akamai Maui Short Course focuses on phenomena associated with the propagation of light and extended light sources/images. The primary goals of the activity include understanding that light travels in straight lines, and that extended light sources/images send out light in all directions from every point. The students investigate phenomena in three stations that illuminate said goals.

Station One and Station Two are investigated in parallel with the students split between them. Station One is a room-sized camera, in which a room is made dark and a small opening is made to a well-lit outer region. In our case we used a door to the outside which was covered in light-blocking fabric with a 1'$\times$1' square cut into it, which is then further covered by a piece of foil to allow variable aperture sizes and shapes to be fashioned. Students stand inside the room see that a inverted image of the outer-region is projected through the aperture onto the walls of the room. By having an instructor pacing back and forth outside of this door while announcing which direction (left or right) they are walking, students are able to realize the horizontal inversion in addition to the obvious vertical inversion. Through their investigations, the students are able to explain the inversion by realizing that light is traveling in straight lines from each point outside, through the aperture, and onto the inside walls of the room. Prompts from facilitators for the students to draw diagrams to help understand what is going on may be necessary as the investigation proceeds. Finally, students who can draw ray diagrams to explain the inversion of the images have proven their understanding of the primary goals of this activity. A common misconception among students is that there is a significant border to the image which is determined in some way by the aperture. This misconception can be worked through via ray-tracing, and critical thinking -- often prompted by the facilitator.

Station Two is referred to as the box-head camera station. In this station, students place cardboard boxes with a small hole in the back and a white ``projection screen'' on the inside front over their heads. If the student is advised to be careful not to block the light's path with their heads and the box is carefully light-insulated (light-blocking fabric draped around the head-hole and the corners taped/sealed off) then they will notice an inverted image of the scene behind them projected onto the front of their box-head camera. Again, the horizontal inversion is slightly more subtle for students to notice and can be highlighted by the instructors wearing shirts/jackets with large text or letters on them. As with the Station One group, prompts from facilitators for students to draw diagrams to make sense of what is causing the inversion are useful as the investigation proceeds. In the end, students must be able to demonstrate (using ray-diagrams or words) an understanding of the cause of the image inversion to prove their understanding of the activity goals.

Station Three is done by all the students at the same time and focuses on an interesting phenomenon which makes use of mirrors of irregular shapes (geometric or random), irregular shaped holes in a piece of card-stock, and direct sunlight -- or some other un-collimated light source with significant physical size. When holding the card-stock or mirrors close to a viewing screen (often the ground or shaded walls) the resulting lit shape takes the shape of the irregular holes or mirrors. However, when the card-stock or mirrors are far from the viewing screen, then the resulting lit shape takes the shape of the light source (typically a disk, after the Sun's image). Students are often confused by this phenomenon at first, thinking that the hole/mirror shape should always determine the resulting light-shape, perhaps reflecting on how shadows take the shape of the object blocking the light. However, given their prior experience with ray-tracing from stations one/two, the students are able to realize that the images of the source get larger and overlap more and more when imaged further from the hole/mirror --  so the image more and more takes the shape of the source. This is demonstrated by the students' verbal explanations and ray-diagrams.

A common misconception that students express is that a certain distance between the image and the hole/mirror brings the image into ``focus''. What they are referring to is actually sharpness of the image sometimes compounded with keeping the distinct components of the scene separated in the image.

Following these investigations, students are given limited time to create and pre\-sent poster presentations describing their process and discoveries. These presentations are immediately followed by a brief synthesis lecture which illustrates and summarizes the main content goals for the activity.

\subsection{Lenses and Refraction}
The Lenses and Refraction (L\&R) activity is designed to enable students to discover for themselves the behavior of light as it passes across medium boundaries. The activity is broken into three stations, each with a dedicated facilitator. The general order of events is 1) Observation of phenomena, 2) Creating and choosing questions to investigate, 3) Using the station equipment to experiment, and 4) Presenting the results and explanations to their peers. A quick introduction given by the activity leader recalls what the students have learned from the previous day's Camera Obscura activity. The students are then permitted to “play around” at the stations and come up with questions about phenomena they cannot yet explain.

\begin{table}[ht]
\begin{center}
\caption{Lenses \& Refraction Timeline}
\smallskip
\begin{tabular}{lc}
\tableline
{\bf Activity Component} & {\bf Time Allotted} \\
\tableline
\tableline
\noalign{\smallskip}
Introduction & 5 min \\
Starters & 20 min \\
Question \& Group Formation & 20 min \\
Focused Investigations & 60 min \\
Lunch Break & 45 min \\
Continue Investigations & 45 min \\
Presentation Preparation & 15 min \\
Poster Presentations & 35 min \\
Synthesis & 30 min \\
\tableline
\end{tabular}
\end{center}
\end{table}

Station One consists of ray boxes and a collection of various lenses of different shapes (i.e., convex, concave, double-convex, etc.). The ray boxes allow for easily distinguishable beams of light to be observed passing through the lenses, which are transparent across all dimensions. Students investigate the reasons why some lenses cause parallel beams from the ray boxes to converge or diverge. The lenses can be combined in series to produce counter-acting effects as well, resulting in the beams returning to their original orientation.

Station Two (and the most popular station by far) is composed of small water tanks and pieces of clear, flat-sided plastic in both common and exotic shapes. The station generates a lot of excitement from the students by using low-power laser pointers as the source of the light beam. Students will see the light beam alternately reflect and transmit when it encounters different mediums at different angles. Questions commonly include the location of the critical angle (where transmission becomes reflection and vice versa), as well as finding ways to “trap” the light beam in the plastic shapes by maximizing the number of internal reflections.

Station Three is often the most baffling for students, and involves the use of lenses to either disperse an incoming image or focus it to a nearby piece of white paper. All the components are arranged on a common track to limit the degrees of freedom, but the apparatus has often still proved very challenging for students to understand. The image is generated by a lamp shining though a cut-out of a cross with unique shapes at each tip to better gauge its orientation when projected through the lenses. Students question what conditions produce a magnified image, a de-magnified image, or no image at all. Due to the complexities involved with this station (i.e., relative distances from focal points and the concept of the ``virtual'' image), it has tended to be the least engaged with by students. However, for those students who do choose to try to understand how images are produced, it provides a comprehensive look at the L\&R content that includes all the concepts from the other two stations.

The presentations given after the investigation have generally conveyed a new, intuitive understanding by the students of Snell's Law and the concept of the critical angle. In addition, the cause of shaped lenses causing light beams to be re-directed is also effectively explained by many of the participants. As previously mentioned, the concepts of the image plane and the cause of magnification of an image are less often investigated, but remain convenient to the facilitators as a challenge for students with more background in optics or for those students who have claimed they already have ``figured out'' the other stations' content.

\subsection{Color and Light}
An understanding of color and light is a fundamental concept in physics.  Pursuit of this understanding also offers an opportunity to further develop process and attitudinal goals.  The list below outlines each of these goals:\\

\noindent Content Goals:
\begin{enumerate}
\vspace{-3mm}
\item[A.] to know that white light is composed of all colors
\vspace{-3mm}
\item[B.] to know that the primary colors of visible light are red, green, and blue
\vspace{-3mm}
\item[C.] to understand additive/subtractive mixing of colors of light
\end{enumerate}
Process Goals:
\begin{enumerate}
\vspace{-3mm}
\item[A.] to raise investigable questions regarding observed phenomena
\vspace{-3mm}
\item[B.] to develop methods/processes to answer a scientific question
\vspace{-3mm}
\item[C.] to produce a clear and organized poster/presentation
\vspace{-3mm}
\item[D.] using predictions to test hypotheses
\end{enumerate}
Attitudinal Goals:
\begin{enumerate}
\vspace{-3mm}
\item[A.] to develop a sense of ownership in an authentic inquiry
\vspace{-3mm}
\item[B.] to have respect for others' ideas
\vspace{-3mm}
\item[C.] to work through frustration in a productive way
\vspace{-3mm}
\item[D.] to work productively with students of different backgrounds (gender, culture, education)
\end{enumerate}

The inquiry consists of three stations at which different phenomena are displayed: subtractive mixing through gel filters, subtractive mixing through dichroic filters, and additive + subtractive mixing through colored shadows.

At Station One, several gel filters of different colors are placed on the flat surface of an overhead projector.  The students notice that when certain color filters overlap, different colors are seen in the overlapping region, sometimes not the ones they expected (e.g., blue and red filters produce black, not purple). The station's primary goal of understanding is that colored materials are not emitting light, but absorbing/transmitting light. As such, the ``red + green + blue=white'' formula is not reproduced by stacking the filters. Rather, the white light of the projector is being selectively subtracted by the filters at certain wavelengths (or colors). A place where students can trip themselves up is by the assumption that the primary colors of light are red, blue, and yellow (as defined by color pigments in the field of visual art). This preconception can be difficult to overcome, and facilitators may need to take extra steps to suggest ways in which the students can disprove this assumption. Indications of success by the students at this station can include drawn diagrams of the filters blocking specific colors of light, while allowing other colors through which correctly combine into the observed, transmitted color.

Station Two consists of three slide projectors (or flashlights) set up so that they display one red, green, or blue (RGB) color each.  When they project onto the same spot from the same distance, they add together to make white light (additive mixing).  When an obstruction such as a ruler is placed in front of the three beams, three distinct cyan, yellow, and magenta (CYM) shadows are produced. Where two shadows overlap, RGB shadows are produced.  When all three shadows overlap, no light transmits (black shadow). The production of ``colored shadows'' usually causes the most confusion among students. This is because they are subconsciously assuming that ``any shadow = no light'', when in fact only one or two light sources are being blocked. When such a problem occurs, facilitators can recall the primary lesson of the Camera Obscura: light travels in straight lines. Success at this station can include a ray-trace of the three lights that shows how the obstruction is preventing one or two colors from reaching a given location on the wall, resulting in a colored shadow. 

Station Three is a set of dichroic filters in front of a white light source.  Students observe that when dichroic filters are placed in front of white light, a portion of the visible spectrum is transmitted while the remaining portion is reflected.  Likewise, when the filters are placed in series, colors transmitted through the upstream filter may be either reflected or transmitted through the downstream filter, depending on the nature of the material from which it is made.  Dichroic filters are named after the color they transmit.  This station is often the most attractive and perplexing to the students, perhaps because it thoroughly explores all three content goals in the face of an interesting piece of engineering (the dichroic filters themselves). Students can become frustrated at this station if they are always using the filters in series, and swapping them around without controlling all the other variables in their set-up. Facilitators should ensure that students try to grasp the one-filter set-up before adding new ones. Success at this station is usually shown by the students correctly determining (with diagrams) which RGB and CYM colors are being reflected or transmitted. The ``thinking tool'' described below has often proved sufficient to give students breakthroughs of understanding at this station.

A thinking tool is used amidst the investigations to stimulate the students' thinking toward the desired direction.  For this, light on the flat surface of an overhead projector is blocked such that only about a 1-cm wide strip of light transmits.  At the center of this strip, three different CYM dichroic filters are placed in parallel, then a curved diffraction grating is fitted at the magnifying portion of the projector.  In this way, the resultant spectrum is dispersed so that each RGB portion is distinct.  Students see that the cyan filter transmits green and blue portions of the spectrum, yellow transmits red and green, and magenta transmits red and blue.  The tool is designed to guide the students toward thinking of CYM colors as being composed of RGB colors, and if they had not achieved content goals A and B yet, they were now given that information so that they can further explore content goal C.

\begin{table}[ht]
\begin{center}
\caption{Color \& Light Timeline}
\smallskip
\begin{tabular}{lc}
\tableline
{\bf Activity Component} & {\bf Time Allotted} \\
\tableline
\tableline
\noalign{\smallskip}
Introduction & 5 min \\
Meta-moment: How to be a good questioner & 5 min \\
Starters & 25 min \\
Break ($+$ question sorting) & 10 min \\
Question \& Group Formation & 10 min \\
Meta-moment: How to work best in groups & 5 min \\
Focused Investigations & 60 min \\
Lunch Break & 45 min \\
Continue Investigations & 30 min \\
Thinking Tool (Demonstration) & 5 min \\
Continue Investigations & 30 min \\
Meta-moment: How to deliver a quality presentation & 10 min \\
Presentation Preparation & 30 min \\
Poster Presentations & 30 min \\
Synthesis & 30 min \\
\tableline
\end{tabular}
\end{center}
\end{table}
To the best of our knowledge, ``meta-moments'' (see Sonnett \& Montgomery, this volume) were first thought of and invoked during this short course. These ``meta-moments'' became key components of this activity.  In earlier activities, we noticed that students were having difficulties performing well at each phase of the standard inquiry process mentioned in the Lenses and Refraction activity section.  We reasoned that this might be due to a lack of understanding why each phase is important.  To address this issue, we decided to implement a series of short meta-cognitive moments where we gave the students a chance to reflect upon each phase.  For example, before the question-generating phase, a facilitator would briefly explain the importance of question-generating then pose a question to the students: ``What makes a good questioner?''.  Students would then respond with answers such as: ``Someone who can pose a diverse group of questions'' or ``Someone who can recognize variables''.  Each of these items would get written for all the students to see as they were mentioned.  If the students had trouble thinking of some items, a prepared list of them could be momentarily called upon to avoid too much of a lull.  These ``meta-moments'' were entirely student-driven so as to give them ownership and a sense of responsibility for fulfilling the phase to the best of their abilities.  They also help guide the students as to how to conduct themselves during each phase and can stimulate significant progress toward attitudinal goals.  Gaining a better understanding of quality performance in each phase alongside reaching attitudinal goals may also make process and content goals easier to achieve. 
     
\subsection{Adaptive Optics Workbench}
The final inquiry activity of the short course is the Adaptive Optics Workbench activity. The core of this activity is a series of three engineering challenge stations designed to teach engineering process skills and optical concepts. An adaptive optics (AO) system is an instrument that is used in an astronomical setting to correct optical distortions of starlight caused by the Earth's atmosphere. Used in a vision science setting, it corrects distortions caused by the tissue of the eye. The AO system used for this activity is outlined in the ``Teaching Optics an Systems Engineering with Adaptive Optics Workbenches'' paper by Harrington et al. in this volume.

\subsubsection{Goals}
There are many optical concepts that are involved in understanding and using an AO system. These include collimation and focusing of light, re-imaging or relays, beam expansion and compression, wavefronts and aberration. Since this system is used at the introductory optics level, the first level content goal is simply to introduce and reinforce focusing, collimation and beam expansion / compression. The AO system can be conceptually broken down into three main sub-systems: distortion creation, distortion sensing and distortion correcting. There is an associated control system (or feedback loop) between the distortion sensor and distortion corrector. Real-world AO systems are quite complex and use expensive components ($\sim$10-30k). By exposing students to a more complex optical system, we can scaffold students' ability to break a complex unfamiliar system into simple interacting components (sub-systems) that can be conceptually outlined. Another implicit goal is to have students apply the knowledge gained in the three previous inquiry activities in a new situation.

The activity is designed to meet the following objectives:
\begin{enumerate}
  \vspace{-3mm}
\item Develop systems thinking by recognizing that a complex tool can be broken down into simpler components that work together as shown by block diagrams.
  \vspace{-3mm}
\item Troubleshoot and diagnose a problem through testing and evaluating multiple solutions.
  \vspace{-3mm}
\item Recognize trade-offs, and how the ``correct'' solution depends upon the requirements of the problem.
  \vspace{-3mm}
\item Become more comfortable in manipulating expensive/complex technology by being required to operate a working AO system.
  \vspace{-3mm}
\item Improve the correction of an image, using the concept of feedback and demonstrating how the wave-front sensor and corrector communicate.
  \vspace{-3mm}
\item Document their methods and work.
  \vspace{-3mm}
\item Characterize both the AO system and the optical phenomena it utilizes to correct distorted images.
  \vspace{-3mm}
\item Communicate their ideas to their peers.
  \vspace{-3mm}
\item Understand how an array of lenses can be used to measure the ``shape'' of an incoming distorted beam.
  \vspace{-3mm}
\item Understand phase conjugation via shaping a mirror to fix a distorted image.
\end{enumerate}

\subsubsection{Activity Outline}
\begin{table}[ht]
\begin{center}
\caption{Adaptive Optics Workbench Timeline}
\smallskip
\begin{tabular}{lc}
\tableline
{\bf Activity Component} & {\bf Time Allotted} \\
\tableline
\tableline
\noalign{\smallskip}
Introduction & 5 min \\
Teaser Talk & 10 min \\
Station 1 Rotation & 50 min \\
Station 2 Rotation & 50 min \\
Station 3 Rotation & 50 min \\
Presentation Planning & 20 min \\
Poster Presentations & 30 min \\
Synthesis & 15 min \\
\tableline
\end{tabular}
\end{center}
\end{table}

First is a teaser talk to introduce the AO Workbench and motivate the activity. The teaser talk is designed to have lots of pictures and videos to describe the motivation for using AO but not to describe the system function. For the station rotation phase, students are organized into three equally-sized groups. The three groups rotate through the three stations spending about 50 minutes at each station investigating. Each station is modeled around the key content for a sub-system or investigates a whole working system. The three groups move through the stations in different orders, so each group's experience of the activity will be slightly different. Students report out on the last station they visit by presenting a poster describing an AO system and how what they learned at their station can be used in understanding the internal functions of the AO workstation.

\subsubsection{Stations}
The first station is distortion sensing and outlines the basic optical concepts behind a traditional ``Shack-Hartmann'' wavefront sensor. The students spend the first part of the activity with graph paper, ray-boxes and acrylic lenses. They are given a simple stated task: ``Measure the relation between how rays hit the lens and where the focal point falls.'' The students then are facilitated toward measuring the focal length of the acrylic lens then mapping out how the incidence angle of the rays relates to the focal point deviation from the optical axis. There is a simple geometric relation that relates where the rays cross to the height of this crossing above the optical axis. Documentation of their measurements is a key facilitation goal as their measurements will be used quantitatively. Once this relation is firmly understood and is somehow tabulated, graphed or functionally described in lab notebooks, the group is given the engineering challenge. A mystery-box has been set up on another table where the light from three ray boxes enters from one side and exits on another side. Inside this box, there are three flat mirrors that bounce the light from three ray boxes at not-quite 45 degrees. This box simulates a situation where some internal optic is broken, mis-aligned or somehow malfunctioning. The students are given the challenge to apply their knowledge of optics to measure the shape of the ``bent mirror'' using only their lenses and without touching the box. This situation can be described as an analogy to the correction of a system without disrupting the system. By using only external instruments, engineers can design a correction without physical access to the instrument. The students need to realize that they can `invert' their previous measurements and make focal-point measurements with their lenses mounted parallel to the box in order to deduce the incidence angle of the beam. From their incidence angles, they can then draw the shape of the ``bent mirror'' if they apply their recently-gained knowledge that the incidence angle is equal to the reflection angle for mirrors (from their previous activities). This focal-point measurement and inversion is conceptually how a wavefront sensor works.

The second station is the distortion correcting station. An optical setup is created with a lamp illuminating a cardboard screen with a cutout letter as a light source. Using simple fresnel lenses in wood holders the light from this object is collimated, reflected, focused and re-collimated, reflected again and finally focused on to a white screen. Conceptually this optical setup can be broken in to the ``sky'' (being collimated light from the object), an aberrator (the first reflection), a telescope (the second lens), and then a re-imaging system (the third lens, second mirror and fourth lens). The first reflection is a flexible mirror that is pre-bent into a cylindrical shape to introduce some optical distortion. This `aberration' very simply mimics what the atmosphere does to starlight just in front of a telescope. The students are shown the screen with both a flat mirror in place of the aberrator (producing a sharp image) and the aberration mirror in place producing a blurred image. The students are given an engineering challenge to ``make the aberrated image as nice as possible''. Students must decide what constitutes ``nice'' and discuss tradeoffs. The students are not allowed to touch the `sky' or the `telescope' but can do anything they like to the re-imaging system. Students apply their knowledge of lenses to draw out a ray-trace of the system and to discuss what might be causing the optical distortion. There are two main methods that can be used to fix the image -- blocking some of the beam at the second reflection (stopping down the system) or replacing the second reflection with a mirror bent to compensate for the aberrator mirror shape (deformable mirror phase-conjugate to the aberration). The students typically measure the focal lengths of the lenses first and then re-align the system. They then investigate the effects of the various materials provided at various points in the beam. They find that a flexible mirror bent certain ways at the right part of the beam will sharpen the image (without dimming it). Conceptually, this is exactly what a deformable mirror does in an adaptive optics system. The solution of blocking some of the beam does make the image sharper, but it also makes it fainter and the trade-offs involved are typically discussed as a potential but undesirable solution.

\begin{figure} [!h, !t, !b]
\begin{center}
\includegraphics[width=0.8\linewidth, angle=0]{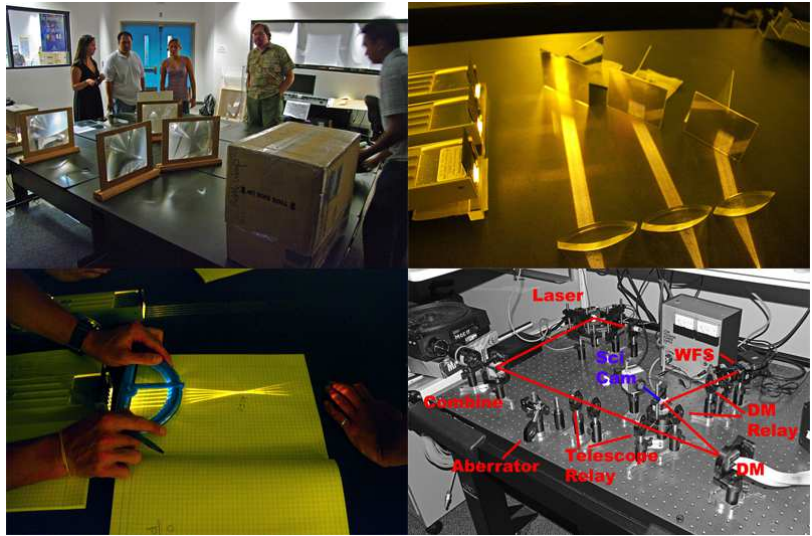}
\caption{The stations for the AO demonstrator activity at the AMSC. The top left image is the wavefront correction station with Fresnel lenses and a bendy mirror. The top right image is the wavefront correction station. The bottom left image shows students deriving the relation between focus location and wavefront tilt. The bottom right image is the systems characterization station using the AO Workbench.}
\end{center}
\end{figure}

The third station is the systems and control-loop investigation using the AO workbench. Students are asked to come up with a thorough written description (words and diagrams) of an AO workstation. The students are first shown the AO system running and given training on using the software. The students are given a task to trace the optical path of the light and to identify what components perform what function. The main goal is the creation of a block diagram and also identifying what optics and detectors correspond to what input / outputs. In order to identify the main function of components and optical groups, the students must trace out the light path. Typically, the students quickly identify the laser which launches light into the system. The beam is large enough that they can also easily identify when the beam is converging or diverging, they can measure beam diameters and they can find focal points. The students are given notecards which block the two detectors.  There are three displays in the system -- one for the computer, one for looking at the corrected image and one for looking at the wavefront sensor camera (which is also fed into the computer). The detector on the wavefront sensor shows an array of little spots formed by the lenslet array. There is an aberrator (a low-powered, long focal length lens) which flips in and out of the system and students can see how this aberrator substantially changes this wavefront sensor image. Since students have practice drawing diagrams of lenses and rays from their previous inquiries, students are facilitated toward drawing a ray-trace for the system. From this ray-trace, they must at least be able to describe beam expansion and compression as well as define collimated beams. Students can also locate the detectors and images in this ray trace, and describe what the output of each detector looks like.

After this ray-trace is performed, the block diagram can be created. The students can see the software running and putting down crosses on the spots on the wavefront sensor image.  As the AO system is turned off and on, they can see the spots move on the screen and can also see the system tracking the spots. The concept usually arises out of group discussions that the computer must somehow be finding and tracking the spots. Since the students can watch the aberrator moving the spots, the concept that the spots somehow can be related to the aberrator shape can be introduced or facilitated. Students also usually notice that there is one mirror which is connected to the computer. Once this is described as a deformable mirror, the concept that the computer tracks the spots and also tells this deformable mirror how to move comes out in students' group discussions.

If the system capabilities, student progress and materials allow, an investigation of the system performance is possible. Certain aberrations can be corrected and certain aberrations cannot. The speed and quality of the correction allows for investigation of control theory and of sub-system communication systems. Using basic materials such as plastic, weak lenses and eyeglasses, students can qualitatively describe what kinds of corrections this system can perform and what corrections it cannot.

The AO workstation has the students work constructively together to build up an understanding of an adaptive optics system thus improving students' group work and communication skills. Students also get to have hands-on experience with fairly sophisticated optical equipment and can build up an understanding of a many-element system using basic optical concepts. In the presentations, students can potentially present a number of concepts and investigation pathways.  Presentations have included many things such as the geometrical relationship between lens focal length, incidence angle and image height, a diagram of the control loop between wavefront sensor, computer and deformable mirror, or how a deformable mirror can be used to ``un-bend'' a light wave by re-imaging the bent aberrator mirror on to another bendable mirror of the same shape (phase conjugation).

\subsection{Communication}
The Akamai program puts a strong emphasis on developing students' communication skills, integrating communication components into all aspects of the program. Students who are accepted into the program are all required to produce a number of deliverables by the end of the eight-week program, all in relation to their own project: an abstract, a ten-minute technical oral presentation, a technical poster, a resume updated with their recent experience, and a personal statement. These assignments, and the support to help all students to be able to produce these deliverables, take place in weekly meetings held in weeks two through seven of the overall Akamai Program. In week one, during the AMSC, we have a set of learning goals that is a preliminary step, or perhaps even a ``warm-up'' for jumping into the more challenging assignments that will come later.

The communication-related goals for AMSC are for students to:
\begin{itemize}
  \vspace{-2mm}
\item Improve their skills in generating scientific explanations
  \vspace{-2mm}
\item Gain confidence in asking questions, proposing investigations or possible solutions, expressing their own ideas, and explaining their findings
  \vspace{-2mm}
\item Gain tools and strategies for working effectively with mentors
  \vspace{-2mm}
\item Write an abstract that includes conventional elements, and is written in S\&E norms and style
  \vspace{-2mm}
\item Demonstrate an understanding of oral communication norms used in S\&E, how they differ from everyday norms of communication, and the value of both
\end{itemize}

A number of components of the AMSC support the achievement of these goals, ranging from formal sessions within the course, to short elements embedded within other activities. Descriptions of the communication elements of the AMSC are below:
\begin{description}
\item[Anatomy of an Abstract:] Students are given a brief introduction to the elements of an abstract such as an introduction, question or problem, methodology, results, and discussion. Students are then grouped into pairs and given two sample abstracts to review. They are provided with a review sheet and asked to discuss with their partner the strengths and weaknesses of the abstract. The abstracts that they are given to review are from past interns (names are not included), so they are very similar to the abstracts that students will be asked to produce on their own project. Once the pairs have reviewed both abstracts, the instructor leads an all-class discussion, making sure that some key points are brought up. For example, novice abstract-writers often write an abstract that is more like an introduction. Or in other cases, the abstract is written such that it is very difficult for the reader to know what the author actually did, versus the work that was done by others that they are building upon. The group also discusses organization, syntax, titles, and other aspects of abstracts.
\item[Abstract Assignment:] Students are given an opportunity to write an abstract on one of the laboratory activities that they have completed in the AMSC, choosing between Camera Obscura and Lenses \& Refraction. Given that both these activities are much like a mini-research experience (a major attribute of ``inquiry'' activities), they make an ideal setting for students to write their first abstract. 
\item[Peer Review of Abstracts:] Students are paired up and review each other's abstract, using the same abstract review sheet used in Anatomy of an Abstract. They are asked to give productive feedback aimed at helping each other improve. All abstracts and reviews are turned in to the instructor, who also reviews and gives feedback.
\item[Informal Poster Presentations:] All inquiry activities include an informal, hand-made poster presentation, where teams of students share with the class the findings of their investigation. Students are asked to state their question, how they went about investigating their question, and what they found. All students on the team are expected to report on some aspect, and students in the audience are encouraged to ask questions. By the end of the course, students have given informal poster presentations multiple times and they are noticeably more comfortable speaking and expressing their ideas.
\item[Communicating with Mentors:] This session is designed to help students learn about what to expect from mentors during their internship experience, how to productively deal with challenging situations, and specific tools for successful communication in the workplace. The session is based on common scenarios that come up, based on years of staff experience in running internships. Students are broken up into small groups and are given a scenario, such as ``you arrive for what you believed was a one-hour meeting with your mentor, but she only has 5 minutes now.'' Each team has a unique scenario, and spends about 5-10 minutes deciding how they will act out an either productive or unproductive way to handle it. If they choose to act out an unproductive way (often quite comical, so very good entertainment), they must also discuss the productive ways of dealing with it. After each scenario, the instructor spends a little time leading a group discussion about the scenario, making sure to cover productive strategies for handling it, and asking other instructors to contribute their experiences. The overall emphasis of this session is that students must be proactive and self-directed -- it is up to them to get the most out of their internship. They should not be too pushy, but should be creative in finding ways to move forward.
\end{description}

\section{Assessment}
\subsection{Assessment of Students}
Much of the student assessment used over the course of the AMSC is ``formative'' assessment -- it is largely collected informally by the facilitators through conversations with the participants in order to better direct and assist the participants' explorations during the inquiry activities. For instance, facilitators ask and look for specific signs that indicate their primary content goals are being reached: during the Camera Obscura and Sun Shadows inquiry, facilitators look for the participants to draw ``ray diagrams'' in which the travel of light is represented as light-rays traveling in straight lines in order to assess whether the students have reached the primary content goals. During the Lenses and Refraction inquiry facilitators ask questions such as ``if the light beam goes into water will it bend more or less steeply?'' in order to assess students' qualitative understanding of Snell's law of refraction. In the Color and Light inquiry facilitators look for students beginning to organize the colors of light into color-wheel style diagrams to demonstrate their understanding of the relationships between the colors of light. During the Adaptive Optics inquiry facilitators look for students to draw box-diagrams to represent the components and relationships between them in the complicated adaptive optics demonstrator system.

One example of summative or ``final'' assessment used during the AMSC is a knowledge survey given at the very beginning and at the very end of the short course. This is given in order to assess the cumulative effect of the AMSC on participants' understandings of the content presented during the inquiry activities. The knowledge survey asked questions requiring conceptual understandings of the various content areas covered. In addition, some ``problem solving'' questions access troubleshooting and logical reasoning skills. The scores and resulting improvement from pre-test to post-test are presented in Table 6. Each student was graded independently by two people, without student identifiers and without knowing which was a pre- or post-survey. We find that on average students improved their scores by 80\%, spread across the regions of the knowledge survey. Thus the students improve their understandings of the content covered as well as their ability to problem solve. Further, because of the length of the course, the students' improvements in understanding are not short-lived. In addition, we note that students \#3 and \#5 had done several of these activities in a course prior to the AMSC. This explains their higher ``pre'' scores, and it is also gratifying to see that they seemed to retain what was learned.

\begin{table}[ht]
\begin{center}
\caption{Pre- and Post-Knowledge Survey Scores and Gains}
\smallskip
\begin{tabular}{|c|ccc|ccc|c|}
\tableline
%{\bf Student \#} & 
%{\bf Pre-Score 1} & {\bf Pre-Score 2} & {\bf Average Pre-Score} &
%{\bf Post-Score 1} & {\bf Post-Score 2} & {\bf Average Post-Score} &
%{\bf Improvement} \\
& {\bf Pre-} & {\bf Pre-} & {\bf Average} &
{\bf Post-} & {\bf Post-} & {\bf Average} & \\
{\bf Student} & 
{\bf Score} & {\bf Score} & {\bf Pre-} &
{\bf Score} & {\bf Score} & {\bf Post-} &
{\bf Improvement} \\
{\bf \#}& {\bf 1} & {\bf 2} & {\bf Score} &
{\bf 1} & {\bf 2} & {\bf Score} & \\
\tableline
1 & 12 & 14 & 13   & 23 & 29 & 26   & 100\% \\
2 & 23 & 26 & 24.5 & 27 & 33 & 30   & 22\% \\
3 & 35 & 39 & 37   & 38 & 40 & 39   & 5\% \\
4 & 12 & 12 & 12   & 29 & 32 & 30.5 & 154\% \\
5 & 29 & 33 & 31   & 31 & 36 & 33.5 & 8\% \\
6 & 19 & 21 & 20   & 37 & 41 & 39   & 95\% \\
7 & 18 & 19 & 18.5 & 26 & 27 & 26.5 & 43\% \\
8 & 30 & 30 & 30   & 42 & 49 & 45.5 & 52\% \\
9 & 17 & 20 & 18.5 & 40 & 43 & 41.5 & 124\% \\
10 & 23 & 29 & 26   & 31 & 34 & 32.5 & 25\% \\
11 & 11 & 10 & 10.5 & 32 & 33 & 32.5 & 210\% \\
12 & 23 & 26 & 24.5 & 39 & 42 & 40.5 & 65\% \\
13 & 19 & 19 & 18.5 & 44 & 42 & 43   & 132\% \\
\tableline
Average & 20.8 & 22.9 & 21.8 & 33.8 & 37.0 & 35.4 & 79.7\% \\
\tableline
\end{tabular}
\end{center}
\end{table}

\subsection{Evaluation of AMSC}
Students are given a survey at the end of the short course, and then again at the end of the eight-week program. Responses are very positive, and highlight the following valuable aspects of AMSC:
\begin{itemize}
\vspace{-2mm}
\item Broadening students' perspective of Science \& Engineering careers
\vspace{-3mm}
\item Practice using reasoning and problem-solving skills
\vspace{-3mm}
\item Communicating, presenting ideas, and interacting with other
\vspace{-3mm}
\item Confidence tackling a problem
\end{itemize}

Evidence of these gains comes from responses to several prompts in both surveys.  At the end of the short course, students complete a short survey on their experience in the AMSC. Results from the 2009 survey indicate that:
\begin{itemize}
\item The overall value of the AMSC was rated highly:
  \begin{itemize}
    \vspace{-3mm}
  \item 75\% (9 of 12) students found the AMSC to overall be ``extremely valuable''
    \vspace{-1mm}
  \item 25\% (3 of 12) found it to be ``very valuable''
    \vspace{-1mm}
  \item 0\% (none) indicated it was ``not at all valuable'', ``somewhat valuable'', or ``valuable''
    \vspace{-1mm}
  \end{itemize}
\item The AMSC changed the way that students were thinking about, or planning for, their career or education. Sample responses to this question include:
  \begin{itemize}
    \vspace{-2mm}
  \item ``I might actually pursue a career in astronomy, I find it really interesting''
  \item ``Not so much education, but my career. I now see that many options exist in Hawaii for high tech jobs, I don't have to suffer on the mainland hopefully''
  \item ``Yes, I like the mechanics as much as the software. I might take up electrical or mechanical engineering''
  \item ``Helped me organize my thoughts about my future through advice from instructors''
  \end{itemize}
\item When asked about how much the AMSC helped them learn about science and engineering processes, students reported that they strongly agreed that the AMSC helped them learn about: 
  \begin{itemize}
    \vspace{-3mm}
  \item Defining or clarifying a question/problem/design (83\% strongly agreed)
    \vspace{-1mm}
  \item Designing their own experiment (100\%)
    \vspace{-1mm}
  \item Designing their own way to solve a problem (100\%)
    \vspace{-1mm}
  \item Using data to develop a scientific explanation or justify a solution (92\%)
    \vspace{-1mm}
  \item Evaluating their design or solution to problem (67\%)
    \vspace{-1mm}
  \item Thinking about or developing alternative explanations/solutions (92\%)
    \vspace{-1mm}
  \item Experience presenting scientific/technical ideas/results (67\%)
    \vspace{-1mm}
  \item Writing abstracts and preparing posters to present results (75\%)
    \vspace{-1mm}
  \item Comments about the above process skills include:
    \begin{itemize}
    \vspace{-1mm}
    \item ``forces you to see a problem differently''
    \item ``I often have problems defining my problem. This has helped''
    \item ``I like to know how things work and figure it out by myself''
    \item ``I loved this. When I make my own experiments I learn a lot''
    \item ``I like that it encourages thinking outside the box''
    \item ``I got to learn to focus on finding the solution instead of just what the right answer is''
    \item ``I have changed my way of thinking''
    \item ``I make a lot of assumptions, this has helped to challenge them''
    \item ``I gained a lot from this. You can solve problems in many ways''
    \end{itemize}
  \end{itemize}
\end{itemize}

Students were also asked to complete a survey at the end of the eight-week program (seven weeks after finishing the AMSC), and reflect back on the value of the AMSC.
\begin{itemize}
\item Students felt that the AMSC prepared them for their internship:
  \begin{itemize}
    \vspace{-3mm}
  \item 54\% (6 of 11) reported that it ``prepared me very well''
    \vspace{-1mm}
  \item 27\% (3 of 11) felt ``adequately prepared''
    \vspace{-1mm}
  \item 18\% (2 of 11) reported that it ``helped somewhat''
    \vspace{-1mm}
  \item Explanations of the above ratings include:
    \begin{itemize}
    \vspace{-1mm}
    \item ``the inquiry prepared me for the work I did at ... Presentation practice also built my confidence to talk in front of people''
    \item ``practice presenting to peers has greatly helped me''
    \item ``learned more on the job, however it prepared me for interacting with my mentor and the thinking skills involved in problem solving''
    \item ``team work, communication, and presentation skills''
    \end{itemize}
  \end{itemize}
\item Students also valued the AMSC in general (regardless of preparation for their internship):
  \begin{itemize}
    \vspace{-3mm}
  \item 45\% rated the AMSC as ``extremely valuable''
    \vspace{-1mm}
  \item 36\% rated it ``very valuable''
    \vspace{-1mm}
  \item 9\% rated it ``valuable''
    \vspace{-1mm}
  \item 9\% rated it ``somewhat valuable''
    \vspace{-1mm}
  \item 0\% rated it ``not valuable''
    \vspace{-1mm}
  \item Comments on the above ratings include:
    \begin{itemize}
    \vspace{-1mm}
    \item ``helped us students how to interact with others, especially with feedback from others''
    \item ``learning the thinking process was valuable''
    \item ``helped me interact well with other people's personalities''
    \item ``presentation skills, communicating ideas''
    \item ``the process of thinking. Question, investigate/experiment, solution''
    \end{itemize}
  \end{itemize}
\end{itemize}

\section{Social and/or Cultural Aspects of the Design in Practice}
There were many opportunities for the design of the AMSC to incorporate and integrate cultural elements. One example is the science and culture educational thread of the short course. As a part of this cultural thread we invited a native Hawaiian elder (Kahu Maxwell) to come and discuss the cultural implications of the growing technological presence on Hawaii with the participants. In addition, we were able to go with Kahu's grandson (Dane) up to the Heiau (sacred place/altar) on the summit of Haleakal\={a} at dawn on the fourth day of the short course where we chanted the traditional Hawaiian song `E Ala \={E}' as the sun rose. From this place of cultural respect, we then toured the technology/telescope sites atop Haleakal\={a}. This science and culture thread gives the participants exposure to and respect for both sides of the necessary balance between cultural heritage and technological innovation. Over the course of their internships Kahu Maxwell worked with each of the participants individually to address the cultural impact and implications of their specific internship project.

During the 2009 AMSC, students were exposed to current debate on scientific development in culturally important areas.  After completing a scientific investigation, students sat in on a community meeting focused on possible means to mitigate the impact of the planned development of a solar telescope on the summit of Haleakal\={a}.  During the meeting, people argued for and against the installation of a new telescope on the summit.  After leaving the meeting, students were graced with a presentation by Hawaiian culture and language advisor Luana Kawa`a.  Her presentation ``Haleakal\={a}: A Sence of Place'' explained the significance of the mountain to the Hawaiian people and was intended give the students some cultural background that they may not have been exposed to.  Later in the week, after visiting the Haleakal\={a} summit in person, students talked in small groups about the pros and cons of telescopes on the mountain.  The discussion allowed students to hear different perspectives from their peers and to articulate their own thoughts on the issue. 

Learners (students, prospective interns) do brief research about the companies they are going to work for in an two hour long activity. They get basic infromation about the history, location, number of employees and areas of interest and service/products of the companies, which are located on Maui (some of them Maui based). This will help prospective interns 1) learn how to perform research about a high tech company and what data to look for and 2) perform optimally in their future internship/employment. 

\section{Considerations for the Future}
The Akamai Maui Short Course is in an ongoing state of evolution.  The course builds upon current research in inquiry teaching and learning and practical experience learned from years past.  There are bound to be many changes as the course is fine tuned in the future.  During the 2009 AMSC the need for modification presented itself when two out of the 12 students (17\%) had previously been exposed to some of the inquiry activities in other courses.  While the AMSC instructors were able to provide these students with fresh challenges within the inquiry topics, the redundancy likely took away from the overall experience.  Due to the small size of Maui's academic community, it is likely that this situation will present itself again in the future.  To better prepare, fresh investigations within each topic should be planned in advance.  In addition, future AMSC instructors should consider diversifying the course to include more engineering related context in the assignments and the assessment of student progress because many AMSC students enter engineering related internships.  By doing so, the added diversity may help instructors avoid the problem of redundancy.

\acknowledgements The Akamai Maui Short Course has evolved and built upon many layers of instructors, input and support. Maui Community College has provided invaluable support and space, year after year. Also many people have been involved in the design and execution of this course beyond the authors represented here. Malika Bell, Scott Seagroves and Nina Arnberg have led a segment within the AMSC on college preparation, job-paths, and communication. Mark Hoffman has led the ``Build a Telescope'' activity. Malika Bell, Hilary O'Bryan and Lani LeBron have provided critical logistical support. Lisa Hunter has overseen, coordinated and ensured the continued success of the AMSC since its beginnings.

In addition, this work has been funded and supported by: National Science Foundation (NSF) Science and Technology Center funding of the Center for Adaptive Optics, managed by the University of California, Santa Cruz, under cooperative agreement \#AST-9876783; NSF \#AST-0836053; NSF \#AST-0850532; NSF \#AST-0710699; NSF/Air Force Office of Scientific Research via NSF \#AST-0710699; the UCSC Institute for Scientist and Engineer Educators; and the University of Hawai`i.
 
Teaching Teams: (and support)
\begin{description}
  \vspace{-2mm}
\item[2003] - Andy Sheinis*, Mark Hoffman, Jenny Patience, Fernando Romero, Juliana Lin, Malika Bell
  \vspace{-2mm}
\item[2004] - Andy Sheinis*, Mark Hoffman, Mike Kuhlen, Sarah Martell, Oscar Azucena, Malika Bell
  \vspace{-2mm}
\item[2005] - Andy Sheinis*, Mark Hoffman, Karrie Gilbert, Mark Ammons, Malika Bell
  \vspace{-2mm}
\item[2006] - Jess Johnson*, Ryan Montgomery, Karrie Gilbert, Eddie Laag, Malika Bell (+ Mark Hoffman)
  \vspace{-2mm}
\item[2007] - Jess Johnson*, Ryan Montgomery, Dave Harrington, Isar Mostafanezhad, Scott Seagroves (+ Mark Hoffman, Hilary O'Bryan)
  \vspace{-2mm}
\item[2008] - Ryan Montgomery*, Dave Harrington, Sarah Sonnett, Isar Mostafanezhad, Mark Pitts, Scott Seagroves (+ Mark Hoffman, Lani LeBron)
  \vspace{-2mm}
\item[2009] - Dave Harrington*, Mark Pitts, Mike Foley, Nina Arnberg (+ Mark Hoffman, Lani LeBron)
  \vspace{-2mm}
\end{description}
(*Lead Instructor)

\bibliography{montgomery}

\end{document}